\documentclass[twoside]{mhd}  
\usepackage{latexsym}
\usepackage{dsfont}
\usepackage{nicefrac}
\usepackage{amsmath}
\usepackage[dvips]{graphicx}

\def\ga{\mathrel{\mathchoice {\vcenter{\offinterlineskip\halign{\hfil
$\displaystyle##$\hfil\cr>\cr\sim\cr}}}
{\vcenter{\offinterlineskip\halign{\hfil$\textstyle##$\hfil\cr
>\cr\sim\cr}}}
{\vcenter{\offinterlineskip\halign{\hfil$\scriptstyle##$\hfil\cr
>\cr\sim\cr}}}
{\vcenter{\offinterlineskip\halign{\hfil$\scriptscriptstyle##$\hfil\cr
>\cr\sim\cr}}}}}
\def\la{\mathrel{\mathchoice {\vcenter{\offinterlineskip\halign{\hfil
$\displaystyle##$\hfil\cr<\cr\sim\cr}}}
{\vcenter{\offinterlineskip\halign{\hfil$\textstyle##$\hfil\cr  
<\cr\sim\cr}}}
{\vcenter{\offinterlineskip\halign{\hfil$\scriptstyle##$\hfil\cr
<\cr\sim\cr}}}
{\vcenter{\offinterlineskip\halign{\hfil$\scriptscriptstyle##$\hfil\cr
<\cr\sim\cr}}}}}

\renewcommand{\vec}[1]{\mbox{\boldmath$#1$}}
\newcommand{\bi}{\begin{itemize}}
\newcommand{\ei}{\end{itemize}}
\newcommand{\be}{\begin{equation}}
\newcommand{\ee}{\end{equation}}
\newcommand{\bea}{\begin{eqnarray}}
\newcommand{\eea}{\end{eqnarray}}
\newcommand{\ds}{\displaystyle}
\mhdhead{0}{0}{1}       % MHD header (vol., issue, first page)
\title{Kinematic simulations of dynamo action with a hybrid
  boundary-element/finite-volume method}        %Paper title
\author{A.~Giesecke, F.~Stefani, G. Gerbeth}    %Author's name 
\institute{Forschungszentrum Dresden -- Rossendorf, Department Magnetohydrodynamics, POB 51 01 19, D -- 01314 Dresden, Germany} 
                                                
\begin{document}

%----------------------- Head
\maketitle
\begin{abstract}
The experimental realization of dynamo excitation as well as theoretical and
numerical examinations of the induction equation have shown the relevance of
boundary conditions for a self-sustaining dynamo.
Within the interior of a field producing domain geometric constraints
or varying material properties (e.g. electrical conductivity of the container
walls or localized high-permeability material) might also play 
a role.
Combining a grid based finite volume approach with the boundary element
method in a hybrid FV-BEM scheme offers the flexibility of a local
discretization with a stringent treatment of insulating magnetic boundary
conditions in almost arbitrary geometries at comparatively low costs. 
Kinematic simulations of dynamo action generated by a well known prescribed mean flow 
demonstrate the reliability of the approach.

Future examinations are intended to understand the behavior of the
VKS-dynamo experiment where the field producing flow is driven by ferrous propellers and
the induction effects of conductivity/permeability inhomogeneities might 
provide the required conditions for the measured dynamo characteristics.
%
%\classification{PACS number(s)}
\end{abstract}
%----------------------- End of Head     

%----------------------- Body       
\section*{Introduction.}
Nowadays, there are nearly no doubts that the mechanism which is responsible
for the generation of astrophysical or planetary magnetic fields is a dynamo
process in which kinetic energy from a suitable flow of a conducting
fluid is transfered into magnetic energy. 
Although the basic idea of this process has already been presented at the
beginning of the 20th century \cite{1919} only few years ago fluid flow
generated dynamo action has been 
realized in the laboratory \cite{2000PhRvL..84.4365G, 2001PhFl...13..561S,
  2007PhRvL..98d4502M}. 
The key parameter that determines the onset of dynamo action is the magnetic
 Reynolds number $\rm{Rm}=\mu_0\sigma {VL}$ where $\sigma$ denotes the
 electrical conductivity, $\mu_0$ the vacuum permeability, ${V}$ a typical
 velocity magnitude 
 and ${L}$ the characteristic size of the considered system. Critical
 values that are necessary to obtain dynamo action in the laboratory are of
 the order 
$\rm{Rm}^{\rm{crit}}\sim 30...100$ which is already technically
demanding.
Therefore, essential efforts are concentrated on possibilities to reduce this
critical value and to increase the 
actual $\rm{Rm}$ of the field producing flow.

From numerical simulations it is known that the boundary conditions and also
boundary layers of stagnant or somehow guided flow could possess supportive as
well as obstructive impacts on the onset of dynamo action
\cite{2005EPJB...47..127A, 2005physics..11149S}.  
In the kinematic regime the backreaction of the field on the flow by the Lorentz
force can be ignored so that
the complexity of the underlying system of equations is significantly reduced 
because only the induction equation with a prescribed velocity field has to be
solved numerically.
Nevertheless, analyzing laboratory experiments 
requires a flexible numerical scheme that
is able to  consider geometric constraints
as well as material properties like conductivity jumps between fluid and
container walls or the high-permeability domains brought in by the iron
propellers that are used to drive the flow in the VKS experiment, at least in
the realization which showed dynamo action up to present \cite{2007PhRvL..98d4502M}.
The finite volume (FV) approach 
provides a fast and robust scheme relying on a local
discretization which delivers
an accurate solution of the kinematic dynamo problem and 
intrinsically maintains the solenoidal character of the magnetic field \cite{2006JCoPh.218...44T}. 
However, a drawback of grid based schemes are the difficulties arising from
non-local boundary conditions as 
they exist in the laboratory in terms of insulating boundaries.
Insulating boundary conditions in non-spherical geometry, in
general, are treated by elaborated schemes, e.g. solving of the Laplace equation in
an extended domain and applying some matching conditions 
 \cite{2003guermond, 2004PhPl...11.2838G, 2006guermond_eccomas}, embedding the
domain in a sphere \cite{1997PhLA..226...75T} or by simplifying approximations
(pseudo vacuum, vanishing tangential field). 
Rather precise results which consider
insulating boundary conditions exactly are provided by the integral equation
approach \cite{2004JCoPh.196..102X, 2004PhRvE..70e6305X}.  
However, the
application possibilities are limited because of enormous computational
resources that are required by this method. 
A different approach that needs less computational power is known as the
boundary element method (BEM) \cite{bem_brebbia}.  
Outside the conducting region, the magnetic flux density $\vec{B}$ is
expressed as the 
gradient of a scalar potential $\vec{B}=-\nabla\varPhi$, where $\varPhi$ is
determined by the Laplace equation: $\Delta \varPhi = 0$.
Making use of Greens second theorem and an appropriate discretization, the
Laplace equation is integrated only on 
the boundary which in the numerical implementation requires the solution of an
algebraic set of equations.
A combined finite volume/boundary element method (FV-BEM) for the induction
equation 
was introduced in \cite{2004JCoPh.197..540I}.
The presented applications were restricted, however, to the decay of an
initial homogenous magnetic field. 
A deeper investigation of the reliability of the numerical approach in case 
of more complex and realistic problems like dynamo action is still missing.

The scope of the present paper is the introduction of the methods and basic
properties of the combined finite 
volume/boundary element method. 
The resulting scheme is adopted for the numerical solution of the kinematic
induction equation in cylindrical coordinates in three 
dimensions with emphasis on the implementation of insulating boundary
conditions. 
The integration of further physical effects like small scale induction effects 
parameterized by an $\alpha$-effect (small scale helical turbulence) are 
easily carried out and
an extension to a scheme that
considers variations and/or jumps in conductivity ($\sigma$) respectively
permeability ($\mu_{\rm{r}}$) is straightforward if corresponding averaging
procedures for $\sigma$ or 
$\mu_{\rm{r}}$ are applied so that the jump
conditions for $\vec{E}$ and $\vec{B}$ at material interfaces are fulfilled
\cite{2001siam...22..1943H}.  
The detailed description of the corresponding methods is beyond the scope of
this publication and will be presented in a subsequent paper.

\section{Equations and numerical methods.}
\subsection{Finite volume method.}
\begin{figure}
\includegraphics[width=12cm]{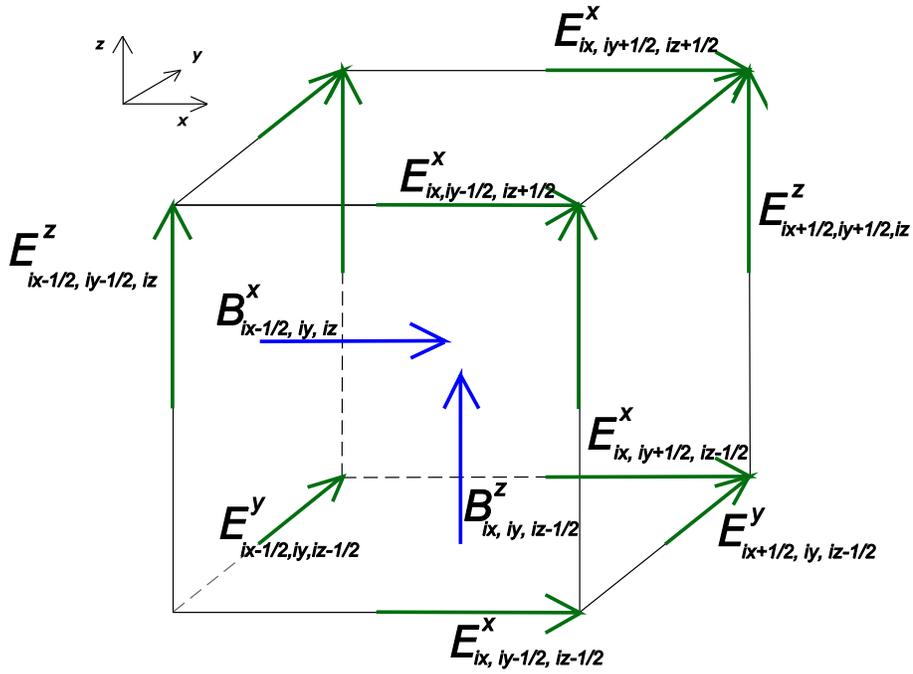}
\caption{Localization of the components of the electric/magnetic field on the
  Cartesian 
  staggered mesh. Not all components are labeled for the reason of clarity.}\label{fig::ct}
\end{figure}

From Faraday's law
${\partial_t\vec{B}}= -\nabla\times \vec{E}$
with the magnetic flux density $\vec{B}$ and the electric field $\vec{E}$ given by
\be
\vec{E}=-\vec{v}\times\vec{B}+\frac{1}{\sigma\mu_0}\nabla\times\frac{\vec{B}}{\mu_r}
\label{eq::efield}
\ee
one immediately retrieves the induction equation 
\begin{equation} 
\frac{\partial\vec{B}}{\partial t}=\nabla\times\left(\vec{v}\times\vec{B}
-\frac{1}{\sigma\mu_0}\nabla\times\frac{\vec{B}}{\mu_{\rm{r}}}\right).
\label{eq::induction}
\end{equation}
Here, $\vec{v}$ denotes the velocity field, $\sigma$ the electric
conductivity, $\mu_0$ the vacuum permeability given by $\mu_0=4\pi\times
10^{-7}\mathrm{VsA^{-1}m^{-1}}$ and $\mu_{\rm{r}}$ the relative permeability,
that describes the ability of the magnetic flux to penetrate a medium.
For most substances like air or non-ferrous conducting materials (copper,
sodium) $\mu_{\rm{r}}$ is very close to $1$ whereas ferrous material
exhibits a relative permeability in the range of $\mu_{\rm{r}}\sim 10^2\dots10^4$.
In the following, only homogenous ($\sigma=\rm{const}$), non-ferromagnetic
($\mu_{\rm{r}}=1$) materials are considered. 
In a finite volume method the computational domain is divided into (small)
control volumes 
where the conservation of all variables is enforced across the
control surfaces (interfaces between neighboring cells). 
Writing the induction equation in conservative form
${\partial_t} \vec{B} + \nabla\times\vec{E} = 0$,
the update of the $x$-component of magnetic field at a timestep $n+1$ in a
Cartesian system is given by
\bea
{B}^{^{x, n+1}}_{_{ix-\frac{1}{2},iy,iz}}&=
{B}^{^{x, n}}_{_{{ix-\frac{1}{2},iy,iz}}}&+\displaystyle\frac{\delta t}{\Delta y}\!
\left(\!E^{^{z,*}}_{_{ix-\frac{1}{2}, iy+\frac{1}{2}, iz}}
-E^{^{z,*}}_{_{ix-\frac{1}{2}, iy-\frac{1}{2}, iz}}\!\right)
\displaystyle\nonumber\\
&&-\frac{\delta t}{\Delta z}\!\!\left(\!E^{^{y,*}}_{_{ix-\frac{1}{2}, iy, iz+\frac{1}{2}}}
-E^{^{y, *}}_{_{ix-\frac{1}{2}, iy, iz-\frac{1}{2}}}\!\right).
\label{eq::fieldupdate}
\eea
In Eq.~(\ref{eq::fieldupdate}) $\vec{E}^{*}$ denotes the electric field at an
intermediate 
time step: 
in an explicit scheme with second order accuracy in time $*$ represents the
time after the update of 
the predictor step whereas in an implicit scheme $*$ represents the actual
timestep $n+1$. 
As indicated by the indices, the localization of the components of
$\vec{E}$ is slightly staggered with regard to the components of $\vec{B}$. 
Fig.~\ref{fig::ct} shows the position of the field components around a single
grid cell labeled $(ix, iy, iz)$.
Vector quantities are defined on the faces ($\vec{B}$, labeled by one
half-integer index) respectively on the edges of a grid cell ($\vec{E}$,
labeled by two half-integer indices) whereas scalar quantities like
conductivity/permeability are defined at the center of 
a grid cell. 
It is convenient to decompose the electric field into an inductive part
$\vec{E}^{\rm{ind}}\propto\vec{v}\times\vec{B}$ and a diffusive part
$\vec{E}^{\rm{diff}}\propto (\mu_0\sigma)^{-1}\nabla\times\vec{B}$.
$\vec{E}_{\rm{ind}}$ is treated explicitly applying the C-MUSCL method introduced in
\cite{2006JCoPh.218...44T}.
Here, only the basics of the scheme are rewritten exemplary for single
components of the involved quantities. 
In a predictor step (at an intermediate timestep $n+\nicefrac{1}{2}$) the
magnetic field on the edges of a grid cell is computed by   
\bea
B_{x, {ix-\frac{1}{2},iy+\frac{1}{2},iz}}^{n+\nicefrac{1}{2},R}\!\!&\!\!\!
=\!\!\!&\!\!\!B^{n}_{x, {ix-\frac{1}{2},iy,iz}}
\!\!+\!\!\left(\frac{\partial B_x}{\partial t}\right)^{n}_{ix-\frac{1}{2},iy,iz}
\!\!\frac{\Delta t}{2}\!+\!\left(\frac{\partial B_x}
{\partial y}\right)^{n}_{ix-\frac{1}{2},iy,iz}\!\!\frac{\Delta y}{2},\label{eq::pred1}\\ 
B_{x, {ix-\frac{1}{2},iy-\frac{1}{2},iz}}^{n+\nicefrac{1}{2},L}\!\!&\!\!\!
=\!\!\!&\!\!\!B^{n}_{x, {ix-\frac{1}{2},iy,iz}}
\!\!+\!\!\left(\frac{\partial B_x}{\partial t}\right)^{n}_{ix-\frac{1}{2},iy,iz}
\!\!\frac{\Delta t}{2}\!-\!\left(\frac{\partial B_x}
{\partial y}\right)^{n}_{ix-\frac{1}{2},iy,iz}\!\!\frac{\Delta y}{2}\label{eq::pred2}, 
\eea
where the time derivative on the right hand side is computed from the known
electric field $\vec{E}^{n}$ at timestep $n$.
The second term on the right hand side $\partial B_x/\partial y$ is
approximated using a monotonized central slope limiter which ensures
positivity preserving, non-oscillating solutions:
\be
\frac{\partial B^x}{\partial
  y}={\rm{minmod}}\left(\frac{B^x_{iy+1}-B^x_{iy-1}}{2\Delta y},
  {\rm{minmod}}\left(2\frac{B^x_{iy+1}-B^x_{iy}}{\Delta y},
2\frac{B^x_{iy}-B^x_{iy-1}}{\Delta y}\right)\right).\label{eq::sl}
\ee
In Eq.~(\ref{eq::sl}) $\rm{minmod}(a, b)$ stands for the minmod limiter
defined by
\be
{\rm{minmod}}(a, b):=\left\{
\begin{array}{rcl}
a & \rm{if} & |a|<|b| {\mbox{ and }} ab > 0\\
b & \rm{if} & |b|<|a| {\mbox{ and }} ab > 0\\
0    & \rm{if} & ab \le 0
\end{array}
\right. 
\ee
The electric field at the intermediate time step $n+\nicefrac{1}{2}$ is then
obtained from the upwind solution of a 2D Riemann problem and is given by:
\bea
E^{{\rm{ind}}, n+\frac{1}{2}}_{x, ix, iy-\frac{1}{2},iz-\frac{1}{2}}&=&
v_y\frac{B^{n+\frac{1}{2},R}_{z,ix,iy-\frac{1}{2},
    iz-\frac{1}{2}}+B^{n+\frac{1}{2},L}_{z,ix,iy-\frac{1}{2},
    iz-\frac{1}{2}}}{2}\nonumber\\ 
&&-v_z\frac{B^{n+\frac{1}{2},R}_{y,ix,iy-\frac{1}{2},
    iz-\frac{1}{2}}+B^{n+\frac{1}{2},L}_{y,ix,iy-\frac{1}{2},
    iz-\frac{1}{2}}}{2}\nonumber\\  
&&-\left|v_y\right|\frac{B^{n+\frac{1}{2},R}_{z,ix,iy-\frac{1}{2},
    iz-\frac{1}{2}}-B^{n+\frac{1}{2},L}_{z,ix,iy-\frac{1}{2},
    iz-\frac{1}{2}}}{2}\nonumber\\ 
&&+\left|v_z\right|\frac{B^{n+\frac{1}{2},R}_{y,ix,iy-\frac{1}{2},
    iz-\frac{1}{2}}-B^{n+\frac{1}{2},L}_{y,ix,iy-\frac{1}{2},
    iz-\frac{1}{2}}}{2}
\eea
where the magnetic field components are the time centered predicted states
interpolated at the edges as determined from Eqs.~(\ref{eq::pred1}) \& (\ref{eq::pred2}).
The final update for the magnetic field is then performed as described by
Eq.~(\ref{eq::fieldupdate}). 
In comparison with a simple scheme where $\vec{v}$ and $\vec{B}$ on the edges 
of a grid cell are computed applying simple arithmetic averages, the C-MUSCL
scheme allows for significant higher magnetic Reynolds numbers without
becoming unstable or exhibiting oscillating solutions.

In an explicit scheme the timestep $\delta t$ is determined by the
Courant-Friedrich-Lax criteria 
\be
\delta t=C\cdot\raisebox{-0.7ex}{${\ds\min} \atop
  {\mbox{\fontsize{9}{9}\selectfont{all cells}}}$}\ds\left(\frac{\Delta
    s_{x, y, z}}{|\vec{v}|},{(\Delta
  s_{x, y, z})^2}{\mu_0\sigma}\right)
\label{eq::timestep}
\ee
with  the minimum of the cell extension in $x, y$ or $z$ direction $\Delta
s_{x,y,z}$ and the Courant factor $C\le 0.5$. 
From expression~(\ref{eq::timestep}) it is immediately evident that  the
timestep is dominated by the 
diffusive part $\propto (\Delta s)^2$ which in a cylindrical system becomes
extremely small for grid 
cells close to the axis.
To relax the constraints of the time step an implicit solver has been
implemented. 
In a first step an intermediate magnetic field $\vec{B}^{*}$ is computed: 
\be
\vec{B}^{*}=\vec{B}^{\rm{exp}}-\delta t\nabla\times\frac{1}{\mu_0\sigma}\nabla\times\vec{B}^{*}
\label{eq::diffus}
\ee
where 
$\vec{B}^{\rm{exp}}$ denotes the magnetic field after the explicit update of
the inductive part as described above and the diffusive part of the electric
field is given in discretized
form by: 
\be
\frac{1}{\sigma\mu_0}\!\left(\nabla\times\vec{B}^{*}\right)_x\!=\!
\frac{1}{\mu_0\sigma}\!\!\left(\frac{B^{y,*}_{ix,iy-\frac{1}{2},iz}\!\!
-\!B^{y,*}_{ix, iy-\frac{1}{2},iz-1}}{\Delta z}
-\frac{B^{z,*}_{ix, iy, iz-\frac{1}{2}}\!\!
-\!B^{z,*}_{ix, iy-1, iz-\frac{1}{2}}}{\Delta y}\right).
\ee
Similar expressions can be written down for the $y$- and $z$-component. 

The resulting algebraic system of equations is solved iteratively for
$\vec{B}^{*}$ using a simple Gauss-Seidel method.  
To remain in the framework of the finite volume scheme the electric
field at time $n+1$ is then computed by 
\be
\vec{E}^{{\rm{diff}}, n+1}=\frac{1}{\sigma\mu_0}\nabla\times{\vec{B}}^{*}
\ee
which is used for the final update of the magnetic field according to
Eq.~(\ref{eq::fieldupdate}).  
\subsection{Treatment of the cylinder axis.}
Although Eq.~(\ref{eq::fieldupdate}) describes the field update in a Cartesian
system, an adoption of the scheme in cylindrical or spherical coordinates is
straightforward (see e.g. \cite{1992ApJS...80..753S, 1992ApJS...80..791S}),
essentially by the application of the appropriate discretization of the operator
$\nabla\times=(\nicefrac{1}{r}\partial_{\varphi}-\partial_z,
\partial_z-\partial_r, \nicefrac{1}{r}\partial_r
(r\cdot)-\nicefrac{1}{r}\partial_{\varphi})$.   
However, at $r=0$ a coordinate singularity exists that prevents the direct
 computation of
 $B^r_{ir=0}$, $B^{\varphi}_{ir=\nicefrac{1}{2}}$ and
 $B^{z}_{ir=\nicefrac{1}{2}}$
so that these quantities have to be treated in a different way.
From the requirement of regularity and uniqueness of the solution at $r=0$
conditions for the behavior of the 
magnetic field at the axis can be derived.
Introducing a decomposition in azimuthal modes 
\be
\vec{B}=\Re(\vec{b}_m(r,z,t)e^{im\varphi}), \quad m=0, 1, 2, 3, ...
\ee
the behavior of the coefficients $b^{r,\varphi,z}_m$ at $r=0$ is determined
by the following -- mode dependent -- relations:
\be
\begin{array}{rrccccl}
\displaystyle
m = 0: & \displaystyle b^r_0 = & b^{\varphi}_0 & = 
& \displaystyle\frac{\partial b^z_0}{\partial r} & = & 0,\\[0.4cm]
\displaystyle
m = 1: & \displaystyle b^z_1 = &\displaystyle\frac{\partial
  b^{\varphi}_1}{\partial r} & 
= & \displaystyle\frac{\partial b^r_1}{\partial r} & = & 0,\\[0.6cm]
\displaystyle
m \ge 2: & b^r_m = & b^{\varphi}_m & = & b^z_m & = & 0.\\
\end{array}
\ee
From these conditions the values of $b^{r, \varphi, z}_m$ at $r=0$ are computed for
every mode $m$ from the values of the corresponding coefficients close to the
axis.
% where
The extrapolation to the axis is based on the radial dependence of $b^{r,
  \varphi, z}_m$ that follows the most general expression for vector
  quantities close to a cylinder axis given by (see
  \cite{2002JCoPh.183..165C}) 
\bea 
B_z(r,\varphi)\!\!&\!\!\!=\!\!&\!\!\!\sum\limits_{m=0}^{\infty}\!r^m
\!\!\left(\sum\limits_{n=0}^{\infty}C^{S}_{mn}r^{2n}\!\right)\!\cos(m\varphi)
\!\!+\!\!\!\sum\limits_{m=0}^{\infty}\!\!r^m
\!\!\left(\sum\limits_{n=0}^{\infty}C^{A}_{mn}r^{2n}\right)\!\sin(m\varphi)\\
B_{r, \varphi}(r,\varphi)\!\!&
\!\!\!=\!\!\!&\!\!\frac{1}{r}\sum\limits_{n=1}^{\infty}C^{\rm{S}}_{0n}r^{2n}
+\sum\limits_{m=1}^{\infty}r^{m-1}
\left(\sum\limits_{n=0}^{\infty}C^{\rm{S}}_{mn}r^{2n}\right)\cos(m\varphi)\nonumber\\
&&+\sum\limits_{m=1}^{\infty}r^{m-1}
\left(\sum\limits_{n=0}^{\infty}C^{\rm{A}}_{mn}r^{2n}\right)\sin(m\varphi).
\label{eq::axis}
\eea
where in the numerical realization
the polynomial expansion in $r$ is truncated at $n=2$.
\section{Boundary element method.}
In case of insulator conditions on the boundary the magnetic field
is computed by the modified integral equation approach presented in \cite{2004JCoPh.197..540I}. 
The method depends on the ability to compute the normal component of the
magnetic field on the boundary by the finite volume scheme and a conceptual
proximity of the finite volume method and the boundary element method
(concerning the location of the components of the magnetic field on a face
centered node). 
The unknown tangential components of the magnetic field at
timestep $(n+1)$ are the
result of a matrix operation on a vector composed of the normal
components of $\vec{B}$ at the surface of the computational domain. 
In the following a sketch of the scheme is given. 

\begin{figure}
\includegraphics[width=10cm,clip=TRUE]{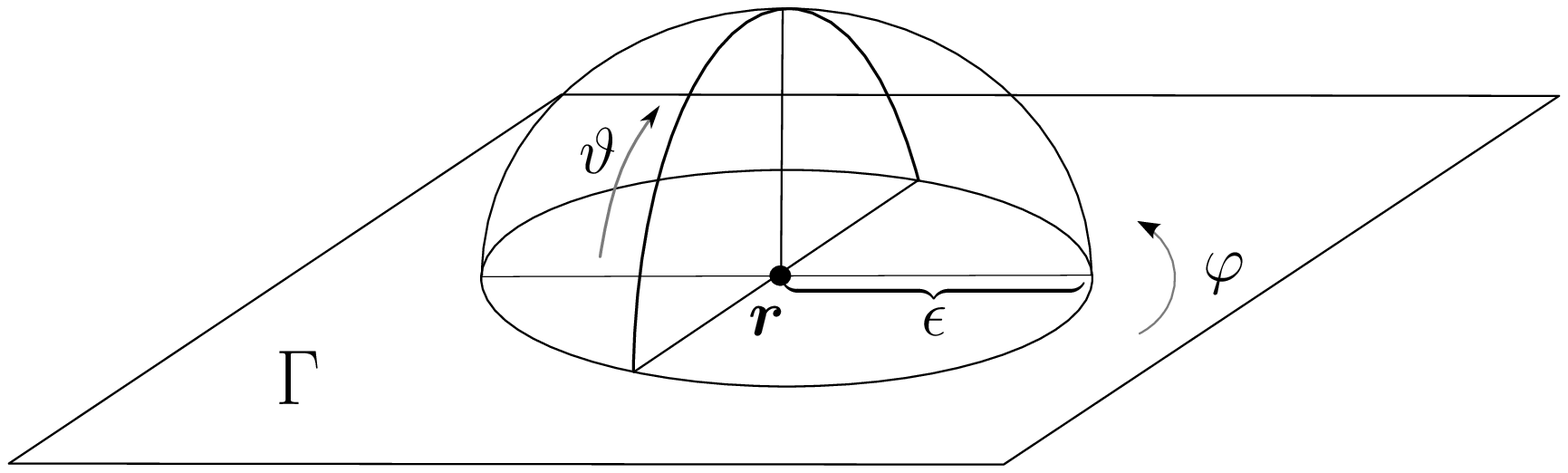}
\caption{Enlarged domain around the source point $\vec{r}$ located on the
  boundary surface $\Gamma$. The boundary integral equation is evaluated  over
an extended domain $\Gamma+\Gamma_{\epsilon}$ in the limit $\epsilon\rightarrow 0$.\label{fig::halfsphere}}
\end{figure}

Insulating domains are characterized by a vanishing current $\vec{j}\propto \nabla\times
  \vec{B}=0$ so that $\vec{B}$ can be expressed as the gradient of a scalar field
$\varPhi$ which fulfills the Laplace equation:
\be
\vec{B}=-\nabla\varPhi \quad\mbox{  with  }\quad \Delta\varPhi =0, \quad
\varPhi \rightarrow O(r^{-2}) \mbox{ for } r\rightarrow\infty.
\label{eq::laplace}
\ee
In a volume $\Omega$ that is bounded by the surface $\Gamma$ Greens second
identity for a scalar function $\varPhi$ and a test- or weighting function $G$
is written as: 
\be
\int\limits_{\Omega}G\Delta \varPhi-\varPhi\Delta G d\Omega
=\int\limits_{\Gamma} G\frac{\partial \varPhi}{\partial
  n}-\varPhi\frac{\partial G}{\partial n}d\Gamma.\label{eq::greens2nd}
\ee
If $\varPhi=0$ the potential is determined by the integral expression 
\be
\varPhi(\vec{r})=\int\limits_{\Gamma} G(\vec{r},\vec{r}')\frac{\partial \varPhi(\vec{r}')}{\partial
  n}-\varPhi(\vec{r}')\frac{\partial G(\vec{r},\vec{r}')}{\partial n}d\Gamma(\vec{r}').\label{eq::inteq}
\ee
where $G(\vec{r}, \vec{r}')$ is called Greens function or fundamental solution
which fulfills 
\be
\Delta G(\vec{r},\vec{r}')=-\delta(\vec{r}-\vec{r}')
\ee
and is given by
\be
G(\vec{r},\vec{r}')=\displaystyle-\frac{1}{4\pi\left|\vec{r}-\vec{r}'\right|}.
\ee
Furthermore, $n$ represents the direction of the normal unit vector on the surface element
$d\Gamma$ and $\nicefrac{\partial}{\partial n}$ is the derivative in the
normal direction: $\nicefrac{\partial}{\partial n}=\vec{n}\cdot\nabla$ so that
$\partial_n\varPhi=-B^{\rm{n}}$ yields the normal component of
$\vec{B}$ on $d\Gamma$.
However, for $\vec{r}\in\Gamma$ Eq.~(\ref{eq::inteq}) is not valid 
since on the boundary $\varPhi$ does not fulfill the H\"older criteria 
($\left|\varPhi(\vec{r})-\varPhi(\vec{r_0})\right|\le Ar^{\alpha}
\mbox{ }\forall\mbox{ }\vec{r}$ with $r\le c$ and $\alpha, c, A > 0$) at every
point $\vec{r}$, 
which is an essential requirement for (\ref{eq::greens2nd}).
The validity of Eq.~(\ref{eq::inteq}) can be extended to points $\vec{r}\in
\Gamma$ if the integration 
domain around a certain source point $\vec{r}$ located on the boundary is
enlarged by a small half sphere with the radius $\epsilon$ and establishing the
limit $\epsilon\rightarrow 0$ (see Fig.~\ref{fig::halfsphere}).

Writing the enlarged domain $\Gamma'=\Gamma+\Gamma_{\epsilon}$ where
$\Gamma_{\epsilon}$ denotes the surface of the half-sphere with radius
$\epsilon$ the first expression on the right side of Eq.~(\ref{eq::inteq}) becomes:
\be
\lim\limits_{\epsilon\rightarrow
  0}\!\int\limits_{\Gamma'}\!\!G(\vec{r},\vec{r}')\frac{\partial
  \varPhi(\vec{r}')}{\partial n}d\Gamma=
\lim\limits_{\epsilon\rightarrow
  0}\!\!\!\int\limits_{\Gamma'-\Gamma_{\epsilon}}\!\!\!\!\!G(\vec{r},\vec{r}')\frac{\partial
  \varPhi(\vec{r}')}{\partial n}d\Gamma+
\lim\limits_{\epsilon\rightarrow
  0}\!\int\limits_{\Gamma_{\epsilon}}\!\!G(\vec{r},\vec{r}')\frac{\partial
  \varPhi(\vec{r}')}{\partial n}d\Gamma.\label{eq::ball}
\ee
With $d\Gamma_{\epsilon}=\epsilon^2\cos\vartheta d\theta d\varphi$
and $\left|\vec{r}-\vec{r}'\right|=\epsilon$
the second expression on the right side of Eq.~(\ref{eq::ball}) vanishes:
\be
\lim\limits_{\epsilon\rightarrow
  0}\int\limits_{\Gamma_{\epsilon}}G(\vec{r},\vec{r}')\frac{\partial
  \varPhi(\vec{r}')}{\partial n}d\Gamma=
 -\lim\limits_{\epsilon\rightarrow
  0}\int\limits_{\Gamma_{\epsilon}}
\frac{1}{4\pi\epsilon}\frac{\partial
  \varPhi(\vec{r}')}{\partial n}\epsilon^2\cos\vartheta d\vartheta d\varphi=0.\label{eq::eps1int}
\ee
In the same manner, the second contribution to the integral expression on the 
RHS of Eq.~(\ref{eq::inteq}) 
is computed as: 
\bea
\lim\limits_{\epsilon\rightarrow
  0}\int\limits_{\Gamma'}\frac{\partial G(\vec{r},\vec{r}')}{\partial
  n}\varPhi(\vec{r}')d\Gamma(\vec{r}')&=&
\lim\limits_{\epsilon\rightarrow
  0}\int\limits_{\Gamma'-\Gamma_{\epsilon}}\frac{\partial G(\vec{r},\vec{r}')}{\partial
  n}\varPhi(\vec{r}')d\Gamma(\vec{r}')\nonumber\\
&+&
\lim\limits_{\epsilon\rightarrow
  0}\int\limits_{\Gamma_{\epsilon}}\frac{\partial G(\vec{r},\vec{r}')}{\partial
  n}\varPhi(\vec{r}')d\Gamma(\vec{r}').
\label{eq::01}
\eea

Unlike in the former case, in Eq.~(\ref{eq::01}),  the contribution of the
integration over the 
$\epsilon$-sphere does not vanish since 
\bea
\lim\limits_{\epsilon\rightarrow
  0}\int\limits_{\Gamma_{\epsilon}}\frac{\partial G(\vec{r},\vec{r}')}{\partial
  n}\varPhi(\vec{r}')d\Gamma(\vec{r}')=
-\lim\limits_{\epsilon\rightarrow
  0}\int\limits_{\Gamma_{\epsilon}}\frac{\vec{n}\cdot(\vec{r}-\vec{r}')}{4\pi\left|\vec{r}-\vec{r}'\right|^3} 
\varPhi(\vec{r}')d\Gamma(\vec{r}')\nonumber\\
=
\lim\limits_{\epsilon\rightarrow
  0}\int\limits_{\Gamma_{\epsilon}}\frac{1}{4\pi\epsilon^2}
\varPhi(\vec{r}') \epsilon^2 \cos\vartheta d\vartheta d\varphi\nonumber\\
=
\int\limits_{\vartheta=-\nicefrac{\pi}{4}}^{\vartheta=+\nicefrac{\pi}{4}}\int\limits_{\varphi=0}^{\varphi=\pi}
\frac{1}{4\pi}\varPhi(\vec{r}') \cos\vartheta d\vartheta
d\varphi=\frac{1}{2}\varPhi(\vec{r}).\label{eq::eps2int}
\eea
Using Eqs.~(\ref{eq::ball}-\ref{eq::eps2int})
the integral equation~(\ref{eq::inteq}) is re-written for $\vec{r}\in\Gamma$ as
\be
\frac{1}{2}\varPhi(\vec{r})=\int\limits_{\Gamma} G(\vec{r},\vec{r}')\underbrace{\frac{\partial \varPhi(\vec{r}')}{\partial
  n}}_{\displaystyle -B^{\rm{n}}(\vec{r}')}-\varPhi(\vec{r}')\frac{\partial
  G(\vec{r},\vec{r}')}{\partial n}d\Gamma(\vec{r}'). \label{eq::bie_phi}
\ee
which is called boundary integral equation.
From Eq.~(\ref{eq::bie_phi}) 
the tangential components of the magnetic field on the boundary
$B^{\rm{t}}=\vec{e}_{\tau}\cdot\vec{B}=-\vec{e}_{\tau}\cdot\nabla\varPhi(\vec{r})$
are computed by:
\be
{B}^{\tau}=2\int\limits_{\Gamma}\vec{e}_{\tau}\cdot\left(\varPhi(\vec{r}')\nabla_{r}\frac{\partial
  G(\vec{r},\vec{r}')}{\partial
  n}+B^{\rm{n}}(\vec{r}')\nabla_{r}G(\vec{r},\vec{r}')\right)d\Gamma(\vec{r}')
\label{eq::bie_b}
\ee
where $\vec{e}_{\tau}$ represents the tangential unit vector on the surface
element $d\Gamma(\vec{r}')$.
Eq.~(\ref{eq::bie_phi}) and ~(\ref{eq::bie_b}) have been derived for a bounded region. 
An infinite volume is treated by introduction of a fictitious surface
$\overline{\Gamma}$ describing a sphere with radius $\overline{R}$ in the limit
$\overline{R}\rightarrow\infty$.
Consider exemplary Eq.~(\ref{eq::bie_phi}) which is re-written including terms
from infinity: 
\bea
\frac{1}{2}\varPhi(\vec{r})=
\int\limits_{\Gamma}G(\vec{r},\vec{r}')\frac{\partial
  \varPhi(\vec{r}')}{\partial n}d\Gamma(\vec{r}')
+\int\limits_{\overline{\Gamma}}G(\vec{r},\vec{r}')\frac{\partial
  \varPhi(\vec{r}')}{\partial n}d\Gamma(\vec{r}')\nonumber\\
-\int\limits_{\Gamma}\varPhi(\vec{r}')\frac{\partial G(\vec{r},
  \vec{r}')}{\partial n}d\Gamma(\vec{r}')
-\int\limits_{\overline{\Gamma}}\varPhi(\vec{r}')\frac{\partial G(\vec{r},
  \vec{r}')}{\partial n}d\Gamma(\vec{r}')\label{eq::infreg}.
\eea
In three dimensions the following asymptotic behavior of the surface element $d\Gamma$
and the fundamental solution $G(\vec{r},\vec{r}')$ occurs for $\overline{R}\rightarrow\infty$:
\begin{subequations}
\bea
d\Gamma(\vec{r}')&=&\left|\vec{\mathcal{J}}\right|d\varphi d\vartheta, \mbox{ }\left|\vec{\mathcal{J}}\right|\sim
O({\overline{R}}^2),\label{eq::asym1}\\
G(\vec{r},\vec{r'})&\sim& O({\overline{R}}^{-1}), \mbox{ }\vec{r}\in\overline{\Gamma},\\
\frac{\partial G(\vec{r},\vec{r}')}{\partial n}&\sim& O({\overline{R}}^{-2}).
\label{eq::asym3}
\eea
\end{subequations}
where $\vec{\mathcal{{J}}}$ denotes the Jacobian.
Combining the regularity condition $\varPhi\propto O({\overline{R}}^{-2})$
  from~(\ref{eq::laplace}) at infinity with the asymptotic behavior given
  in~(\ref{eq::asym1}--\ref{eq::asym3}) it is ensured that the
integral expressions  in~(\ref{eq::infreg}) that involve $\overline{R}$ vanish
  for ${\overline{R}}\rightarrow\infty$. 

\begin{figure}
\includegraphics[width=12cm,clip=TRUE]{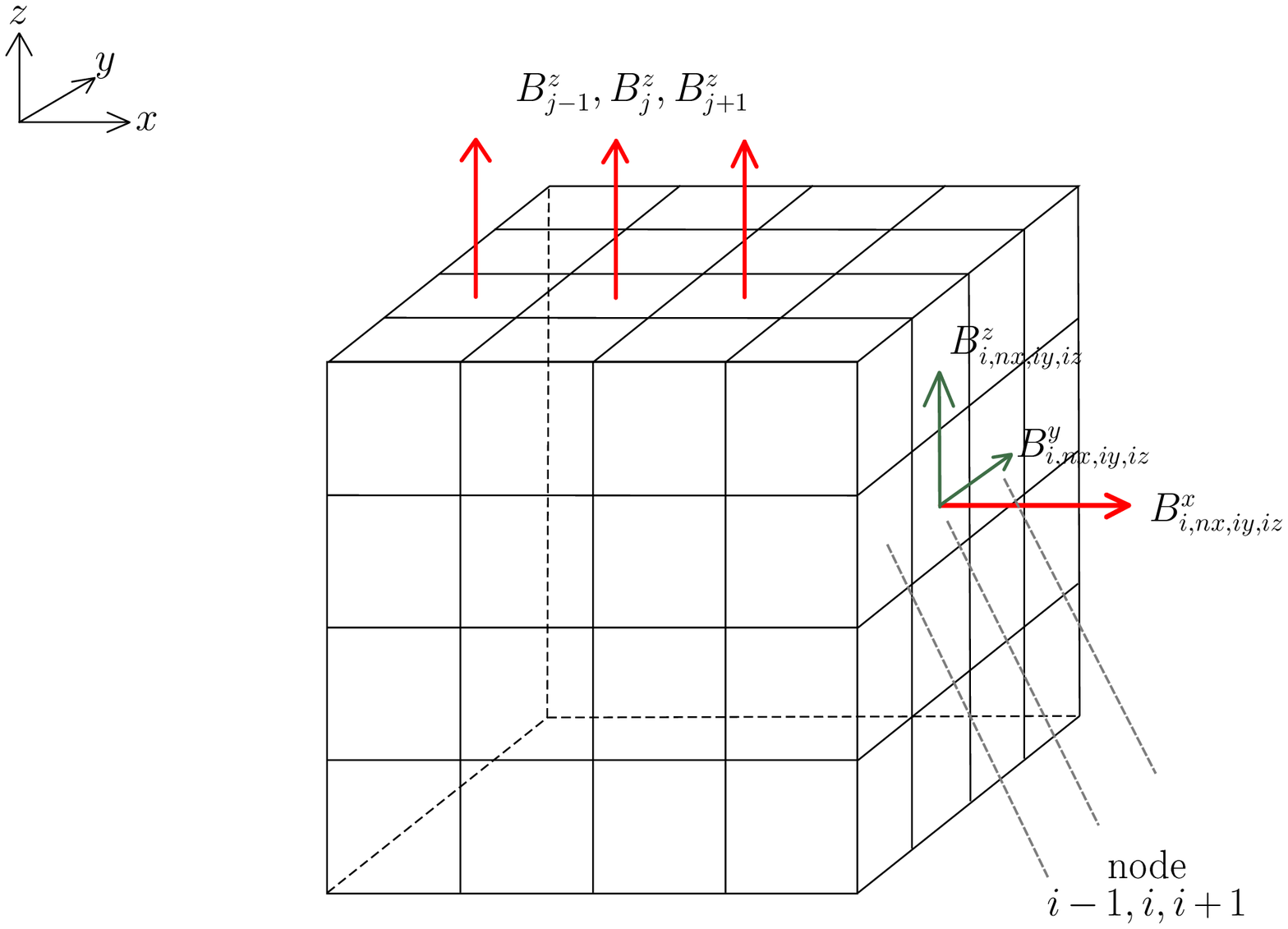}
\caption{Grid structure and boundary element discretization. $i, j$ denote a
  global ordering number.}\label{fig::bem}
\end{figure}

A discretization of the system ~(\ref{eq::bie_phi}) and~(\ref{eq::bie_b}) yields an
algebraic system of equations which allows the computation of the (unknown)
tangential components of the magnetic field.
The natural way to define the boundary elements is an application of the
tessellation provided by the finite volume discretization on the domain
surface,
where every element has one face centered node at which the normal field
component $B^{\rm{n}}$ is located (see Fig.~\ref{fig::bem}). 

After the subdivision of the surface
$\Gamma$ in "small" boundary elements $\Gamma_j$ with $\Gamma=\cup\Gamma_j$
the potential $\varPhi_i=\varPhi(\vec{r_i})$ and 
the tangential field
${B}^{\rm{t}}_i={B}^{\rm{t}}(\vec{r}_i)=-\vec{e}_{\tau}\cdot(\nabla\varPhi_i)$
are given by  
\begin{eqnarray}
\frac{1}{2}\varPhi_i&=&-\sum\limits_j
{\underbrace{\left({\int\limits_{\Gamma_j}\frac{\partial G}{\partial
    n}(\vec{r}_i,\vec{r}')\rm{d}\Gamma_j'}\right)}_{\displaystyle
    \mathcal{H}^1_{ij}}}\varPhi_j
-{\sum\limits_j}{\underbrace{\left({\int\limits_{\Gamma_j}G(\vec{r}_i,\vec{r}')\rm{d}\Gamma_j'}\right)}_{\displaystyle
  \mathcal{H}^2_{ij}}}B^{\rm{n}}_j\nonumber
\\[-0.5cm]
&&\label{eq::discretebem}
\\[-0cm]
B^{\rm{t}}_i&=&\sum\limits_j{\underbrace{\left({\int\limits_{\Gamma_j}2\hat{\vec{e}}_{\tau}\cdot\nabla_{\!r}\frac{\partial
  G}{\partial n}(\vec{r}_i,\vec{r}')\rm{d}\Gamma_j'}\right)}_{\displaystyle
    \mathcal{H}^3_{ij}}}\varPhi_j
+\sum\limits_j{{\underbrace{\left({\int\limits_{\Gamma_j}2\hat{\vec{e}}_{\tau}
\cdot\nabla_{\!r}G(\vec{r}_i,\vec{r}')\rm{d}\Gamma_j'}\right)}
_{\displaystyle\mathcal{H}^4_{ij}}}} B^{n}_j\nonumber.
\end{eqnarray}
Eq.~(\ref{eq::discretebem}) introduces a global ordering of the quantities
$\varPhi$ and ${B}^{\rm{t}, \rm{n}}$ 
defined by an explicit mapping of the grid-cell
indices $(ix, iy, iz)$ on a global index $(i)$ with $i=0,1,2,\cdots,N$
where $N=2\cdot(nz\cdot ny+nz\cdot nx+ny\cdot nx)$ represents the total
number of boundary elements. 
Then $\varPhi$, ${B}^{\rm{t}, \rm{n}}$ can be considered as large vectors and
abbreviating the integral expressions in
(\ref{eq::discretebem}) with $\mathcal{H}^k (k=1{,}...{,}4)$ the system can be
re-written in a matrix representation:
\bea
\frac{1}{2}\varPhi_i&=&-\mathcal{H}^1_{ij}\varPhi_j-\mathcal{H}^2_{ij}B^{\rm{n}}_j,
\\[0.2cm]
B^{\tau}_i&=&\mathcal{H}^3_{ij}\varPhi_j+\mathcal{H}^4_{ij}B^{\rm{n}}_j.
\eea
Finally, a  linear, non-local expression for the tangential field components in
terms of the normal components results:
\begin{equation}
\vec{B}^{\rm{t}}=\left(\mathcal{H}^3\otimes\left(\frac{1}{2}\cdot\mathds{1}-\mathcal{H}^1\right)^{-1}\otimes
\mathcal{H}^2+\mathcal{H}^4\right)\otimes \vec{B}^{\rm{n}}=\mathcal{M}\otimes\vec{B}^{\rm{n}}.
\end{equation}

The numerical computation of the matrix elements $\mathcal{H}^k_{ij}$ is performed applying a
standard 2D-Gauss-Legendre Quadrature method. 
However, for $i=j$ the integral expressions in Eq.~(\ref{eq::discretebem}) become
singular so that $\mathcal{H}^k_{ii}$ have to be treated separately.
Since ${B}^{\rm{t}}$ is computed from the derivative of $\varPhi$ the scalar
potential is only determined except an additional constant.
Fixing this constant results in a relation between diagonal and off-diagonal
elements of the matrices $\mathcal{H}^1$ and $\mathcal{H}^3$ (details see \cite{2004JCoPh.197..540I}): 
\begin{equation}
\mathcal{H}^1_{ii}=-\frac{1}{2}-\sum\limits_{j\ne i}\mathcal{H}^1_{ij}\quad\mbox{ and
}\quad 
\mathcal{H}^3_{ii}=-\sum\limits_{j\ne i}\mathcal{H}^3_{ij}.
\end{equation}

The matrix elements $\mathcal{H}^2_{ii}$ are weakly singular and are computed
numerically without further difficulties after applying a
special cubic coordinate transformation where the Jacobian of the
transformation has a minimum at the singularity \cite{1987__telles}.

The integral expression that determines the diagonal element of $\mathcal{H}^4$ exhibits
  a strong singularity with a vanishing Cauchy principal value so that
  $\mathcal{H}^4_{ii}$ can be computed by excluding some small
  $\epsilon$-vicinity around $\vec{r}_i$. 
Since the elements of $\mathcal{H}^k$ only depend on the
geometry and the discretization of the problem the computation of
$\mathcal{M}$ has to be 
carried out only once. 
However, since the computation of $\vec{B}^{\rm{t}}=\mathcal{M}\otimes
\vec{B}^{\rm{n}}$ 
requires a matrix multiplication with a matrix $\mathcal{M}$ of size
$(2N\times N)$ a large 
amount of memory is required for $\mathcal{M}$ which limits the maximal 
achievable resolution. 
This
restriction is slightly less severe in cylindrical coordinates where the
periodicity in azimuthal direction reduces the necessary size of the matrix
$\mathcal{M}$.  
\section{Results.}
\subsection{Simple test case.}
\begin{figure}
\includegraphics[width=16cm]{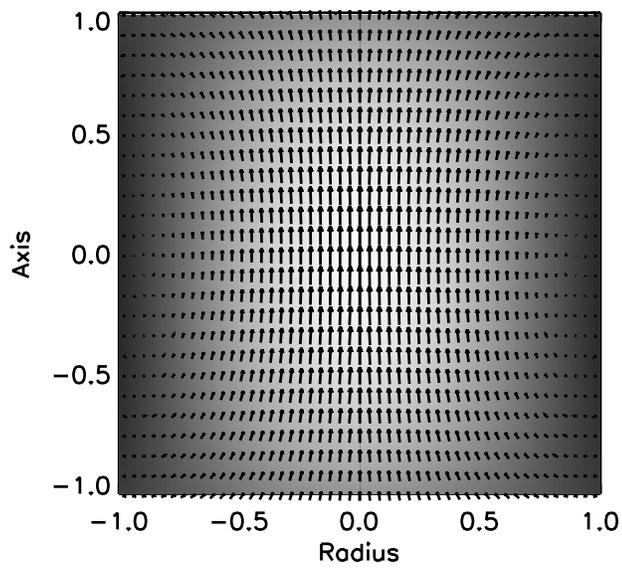}
\caption{
Free decay of an initially random magnetic field. Snapshot of the field structure after one diffusion time. Note the
non-vanishing contributions of the tangential components at the boundary
caused by the insulating boundary conditions.
}\label{fig::dipole}
\end{figure}

\begin{figure}
\includegraphics[width=10cm]{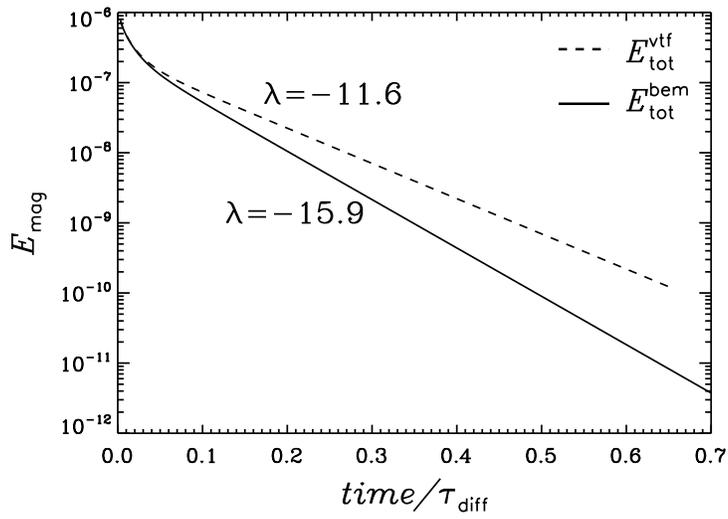}
\caption{Growth rate of a freely decaying magnetic field in a cylinder. 
The solid (dashed) curve shows the total energy applying insulating
(vanishing tangential field) boundary conditions.
}
\label{fig::dipole_decay}
\end{figure}

The free decay of a magnetic field is a simple
test problem where only the diffusive part of the induction equation
($\partial_t \vec{B}\propto\Delta\vec{B}$) is considered.
A (small) challenge for the scheme arises by initially randomly
 distributed field components which involves the presence of all (resolvable)
 modes.
The higher modes decay rather fast and the final solution is dominated by the
axisymmetric dipole mode 
which corresponds to the eigenfunction of the system with the lowest
eigenvalue.  
Fig.~\ref{fig::dipole} shows the structure of the decaying field in a
cylinder of height $H=2$ and radius $R=1$ after approximately one diffusion 
time ($\tau_{\rm{diff}}=\mu_0\sigma
 R^2$) which is dominated by an axial dipole.
The temporal behavior of the total magnetic energy 
$E_{\rm{mag}}=({2\mu_0})^{-1}\int\vec{B}^2dV$
is shown in Fig.~\ref{fig::dipole_decay}.
After the system settles down to its eigensolution (at $t\sim
 0.1\tau_{\rm{diff}}$) a simple exponential decay $\propto 
 e^{-\lambda {t}/{\tau_{\rm{diff}}}}$ is observed.

The influence of the boundary conditions is obvious in comparison
with the decay rate in case of {\it{vanishing tangential field}} boundary
conditions (VTF, 
dashed curve) where a significant slower decay takes place.  
Table~\ref{tab::decay_rates} shows the decay rate for both types of boundary
conditions in comparison with the results obtained by
\cite{2005physics..11149S} applying the 
integral equation approach (IEA) and a differential equation approach (DEA)
where the Laplace equation is solved in the exterior. 
The decay rate of the $m=0$ mode in case of insulating boundaries is larger
than for VTF conditions because of smaller or even vanishing field
gradients in the latter case.
The decay rate for the $m=1$ mode achieved from the FV-BEM scheme is
approximately 4\% larger than the decay rate obtained from the IEA. 
The observed deviation between the different schemes might arise through a
couple of reasons which will be discussed in the following subsection.
\subsection{Von-K\'arm\'an flow.}
The experimental realization of a von-K\'arm\'an-like flow in the VKS
experiment is driven by two counter-rotating propellers located close to the
end-plates of the cylindrical vessel. 
An analytic prescribed velocity field that roughly resembles the mean flow in
the VKS experiment is given by the so called MND-flow
(Mari\'e-Normand-Daviaud, \cite{2004phfl}):
\begin{eqnarray}
v_r&=&-0.5\pi\cos\!\left({\pi z}\right)r(1-r)^2(1+2r)\nonumber\\
v_{\varphi}&=&4\epsilon r(1-r)\sin\left({0.5\pi z}\right)\label{eq::s2t2}\\
v_z&=&(1-r)(1+r-5r^2)\sin\left({\pi z}\right)\nonumber
\end{eqnarray}
where $\epsilon$ describes the relation between toroidal and poloidal component of
the velocity (here given by $\epsilon=0.7259$ which has turned out to be an optimum
value for generating a dynamo \cite{2004phfl}). 

\begin{figure}
\includegraphics[width=14cm, angle=0]{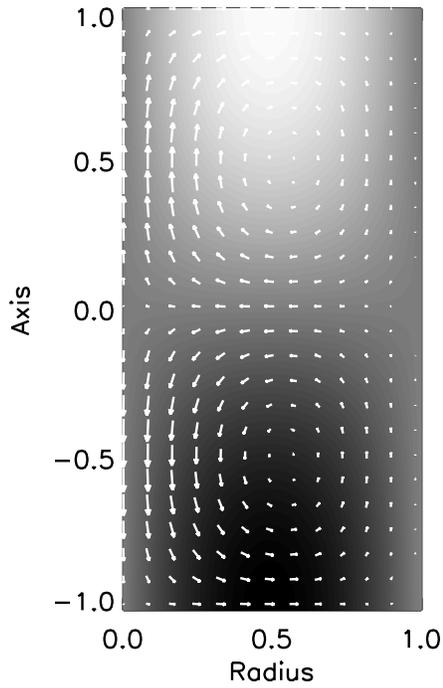}
\caption{Toroidal (shaded contours) and poloidal (arrows) components of the
  von-K\'arm\'an-like flow prescribed by Eq.~(\ref{eq::s2t2}).}\label{fig::flowpattern}
\end{figure}

\begin{figure}
\includegraphics[width=6.2cm,clip=TRUE,angle=0]{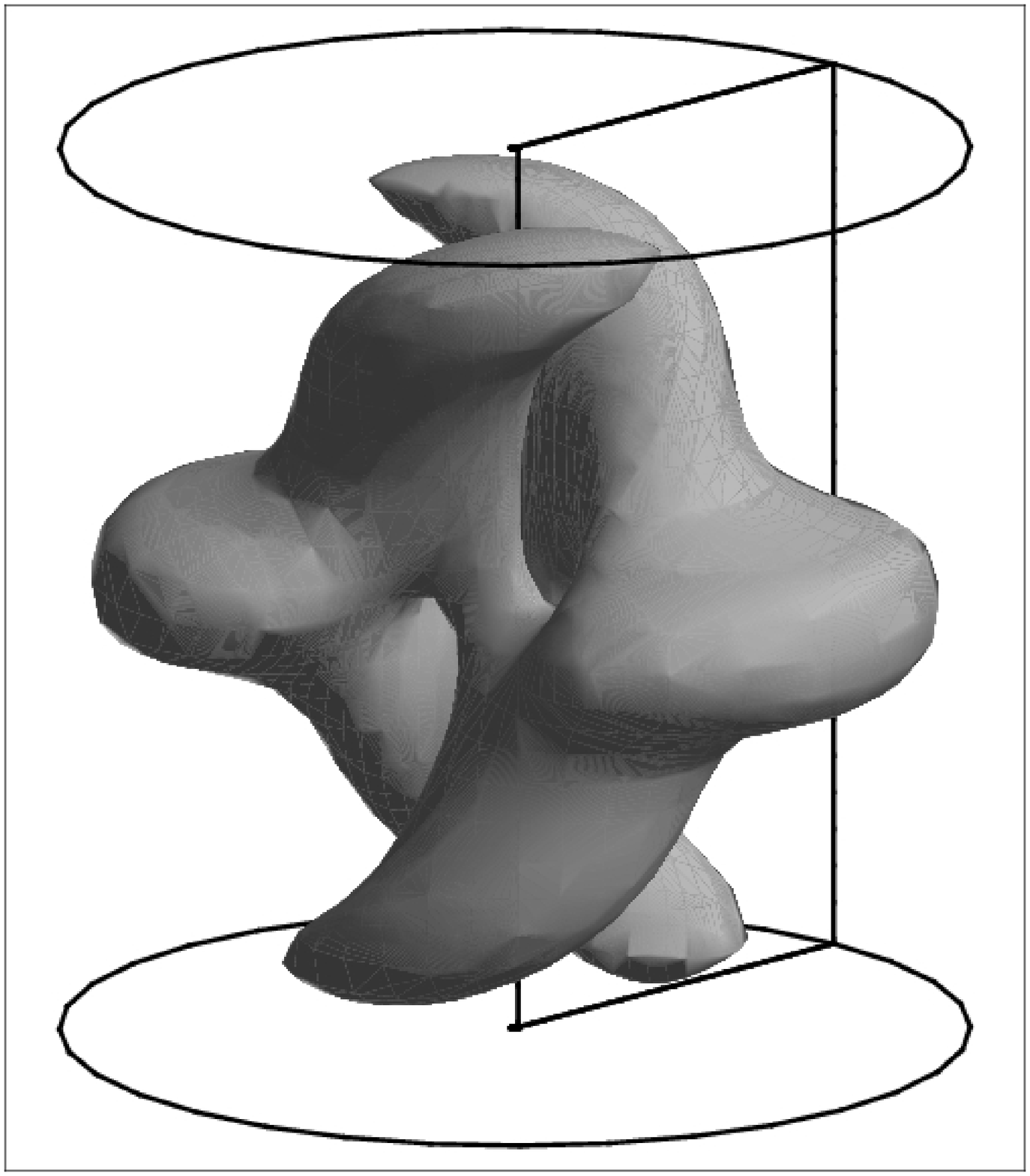}
\nolinebreak[4!]
\includegraphics[width=6.2cm,clip=TRUE,angle=0]{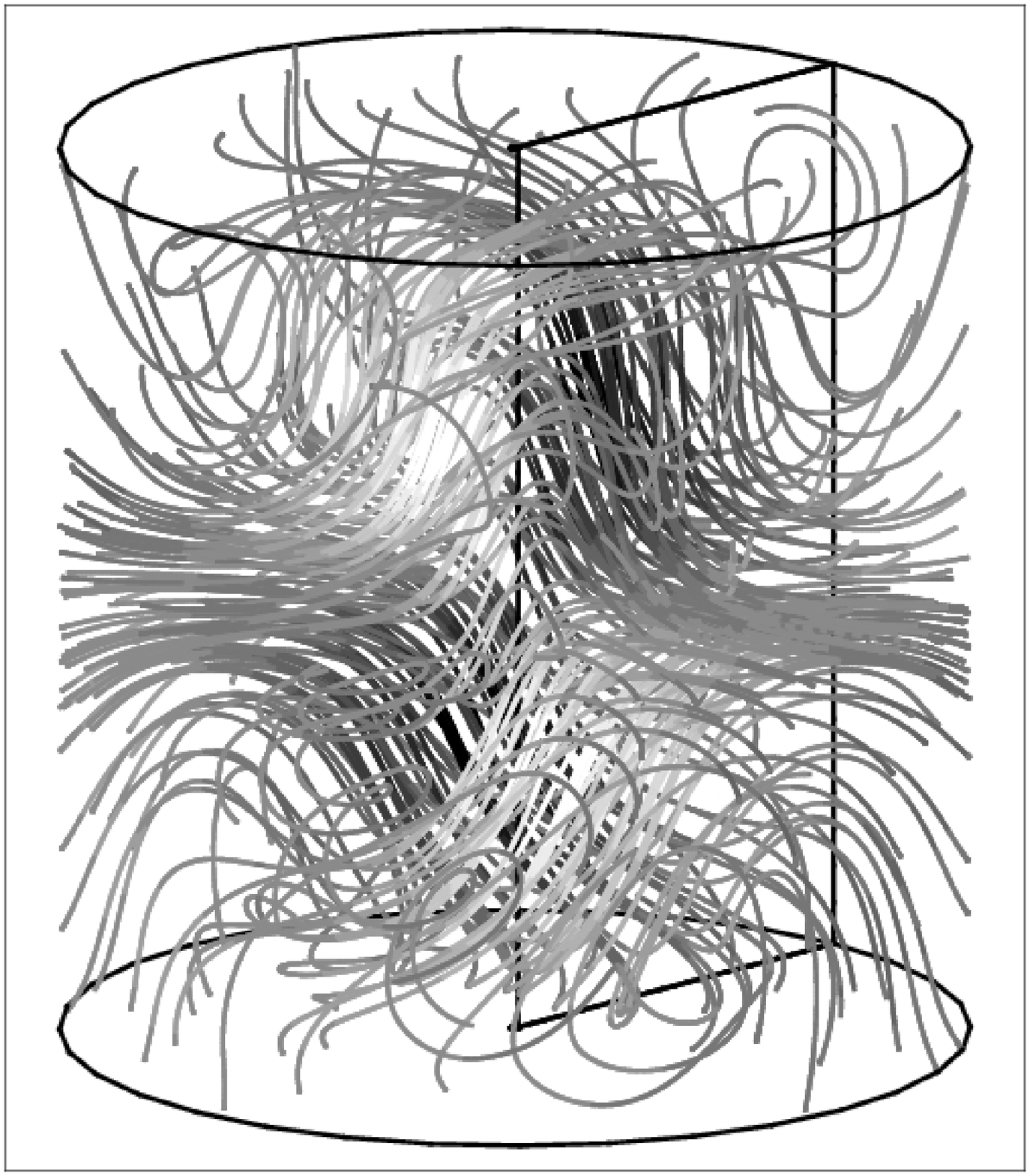}
\caption{Left panel: Isosurface of the magnetic energy at 20\% of the maximum
  value. Right panel: time snapshot of the magnetic field structure represented by
  the streamlines. $\rm{Rm}=80$.}\label{fig::vks-field}
\end{figure}

\begin{figure}
\includegraphics[width=10cm]{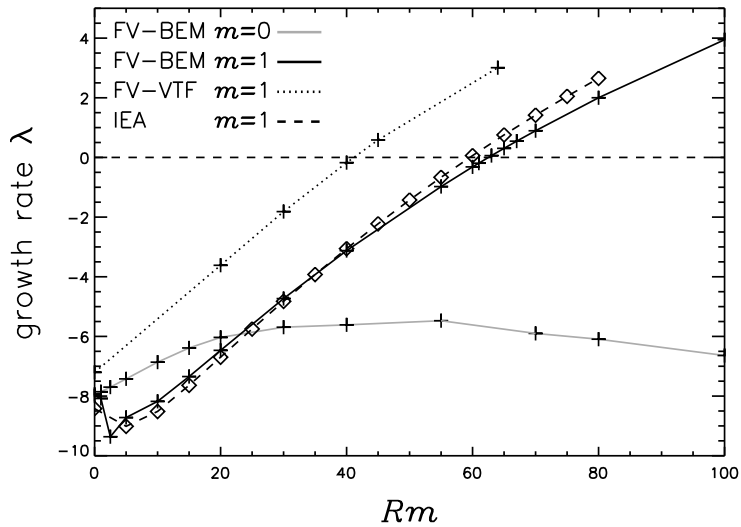}
\caption{Growth rate of the dominating modes ($m=0$, grey solid curve and $m=1$,
  black solid curve) in dependence of the magnetic Reynolds number $\rm{Rm}$. The black dashed line corresponds to the growth
  rates of the $m=1$ mode obtained from the IEA
  \cite{2005physics..11149S}. The dotted curve shows the results for vanishing
  tangential field conditions (VTF).}\label{fig::vks_grothrate_mode} 
\end{figure}

The structure of this flow is essentially dominated by two toroidal and two
poloidal cells which are visualized in Fig.~\ref{fig::flowpattern} where the
grey scaled contours represent the azimuthal flow component and the arrows
represent the poloidal flow components.
If the Reynolds number -- defined as ${\rm{Rm}}=\sigma\mu_0 R
V_{\rm{max}}$ -- exceeds a critical value, dynamo action takes place.
Detailed examinations of the induction effects of the MND-flow -- in particular on
the effects of side and lid layers containing a
stagnant fluid component that surrounds the active region -- can be found in
\cite{2005physics..11149S}.
Here, only simulations without outer fluid layers are discussed. 
The initial condition is again given by a divergence-free random field, which
-- if the applied magnetic Reynolds number exceeds the
critical value -- after a timespan of about a diffusion
time passes into a growing smooth field (Fig.~\ref{fig::vks-field}) field.
The obtained field structure essentially represents the well known banana-cell like
behavior (\cite{2005physics..11149S, 2005PhFl...17k7104R}) corresponding to a
field dominated by the $m=1$ mode  (see left hand side of Fig.~\ref{fig::vks-field}).  
The right hand side of Fig.~\ref{fig::vks-field} presents a snapshot of the
streamlines of the magnetic field from a typical simulation ($\rm{Rm}=80$)
from which the clear dominance of the equatorial oriented dipole solution is apparent.
A quantitative comparison between the FV-BEM and the IEA schemes has been done by means of the
growth-, respectively decay rates for the $m=1$ mode.
If ${\rm{Rm}} \la {\rm{Rm}}^{\rm{crit}}$ the field decays and the slowest decaying eigenmode can
be observed. 
For the 3D grid based FV-BEM scheme the growth rates have been computed from
the temporal behavior of the volume averaged
Fourier amplitudes (azimuthal decomposition) of the axial field component\footnote{All components of the magnetic
  field behave similar, however, $B_z$ is the dominant component.}. 
The growth rates of the dynamo (or decaying) state in dependence of the applied
$\rm{Rm}$ are presented in Fig.~\ref{fig::vks_grothrate_mode}. 

Both schemes provide rather similar results for small and moderate
Reynolds numbers.
Small but systematic deviations between the IEA and the FV-BEM occur for $\rm{Rm}
\ga 45$.
The main and most probable reason for this characteristic behavior might be
the rather low resolution attainable by the IEA which is restricted to
$20\times 20$ points in a 2D simulation due to the enormous computational power required
by this method.
A further influence might be obtained from the coupling of different field
modes that necessarily occurs in the 3D FV-BEM scheme whereas the IEA only
considers one single mode. 
However, it seems unlikely that this influence has a remarkable effect since
the amplitudes of the non-dominating modes are always suppressed by several
orders of magnitude.

For the FV-BEM scheme the critical Reynolds number is obtained from
interpolation of the growth rates around the occurrence of dynamo action
and is given by ${\rm{Rm}}^{\rm{crit}}=62.6$.
This value is situated slightly above the value
reported in \cite{2005PhFl...17k7104R} ($\rm{Rm}^{\rm{crit}}=58$) or the
results of simulations with the DEA
($\rm{Rm}^{\rm{crit}}=61.5$) respectively the IEA
($\rm{Rm}^{\rm{crit}}=59.6$) reported in \cite{2005physics..11149S}.
For idealizing VTF conditions the critical Reynolds number is computed as
$\rm{Rm}^{\rm{crit}}=41.2$ (see dotted curve in Fig.~\ref{fig::vks_grothrate_mode}. 
The significant deviation between the critical Reynolds number for physical boundaries and for
idealizing (VTF) boundaries depends on the geometry of the system and
becomes smaller for a larger relation between height and radius. 
Beside ${\rm{Rm}}^{\rm{crit}}$, a second distinguished point exists, at which the
transition of a $m=0$ dominated solution to a $m=1$ dominated solution
proceeds, which occurs at $\rm{Rm}\approx 25$. Below this value the axial
dipole dominates the field structure, however, the $m=0$ mode never becomes
unstable to dynamo action.
Note, that for small Reynolds numbers both schemes yield a slight minimum for the decay
rate of the $m=1$ mode, so that for ${\rm{Rm}}\la 15$ this mode decays faster than
  without any flow.

\section{Conclusion.}
A fast and easy to handle code base for 3D simulations of the kinematic induction
equation has been developed which is able to treat insulating boundary
conditions in a stringent way.
The reliability of the approach has been tested against known results of
dynamo action generated by the (analytic prescribed) MND-flow and a good agreement was achieved
in comparison with the results of the integral equation approach.
The IEA is also able to treat insulating boundary conditions but requires much more
computational power and provides less flexibility concerning the addition of
further physical effects or geometric constraints.

Significant differences for the critical Reynolds number and the growth rates
exist between simplifying boundary conditions and insulating boundaries. These
discrepancies become smaller for higher aspect ratios.
However, in case of a realistic relation between height and radius as it
usually is realized in the laboratory, a thorough consideration of the
appropriate boundary conditions is indispensable.

The flexibility of the scheme facilitates the addition of further physical
terms like an $\alpha$-effect as additional dynamo source or the consideration
of temporal fluctuations of the mean flow.
A direct extension of the scheme which will be presented in a subsequent paper
will be able to consider conductivity/permeability inhomogeneities.
\Thanks{Financial support from Deutsche Forschungsgemeinschaft (DFG) in frame of the
Collaborative Research Center (SFB) 609 is gratefully acknowledged.}

%-------------------- End of Body
\bibliographystyle{mhd}
\bibliography{references}

\begin{thebibliography}{10}

\bibitem{1919}
{\sc J.~{Larmor}}.
\newblock How could a rotating body such as the sun become a magnet?
\newblock {\it {\it{Rep. Br. Assoc. Adv. Sci. A}}\/},  (1919), pp.~159--160.

\bibitem{2000PhRvL..84.4365G}
{\sc A.~{Gailitis}, et~al.}
\newblock {Detection of a Flow Induced Magnetic Field Eigenmode in the Riga
  Dynamo Facility}.
\newblock {\it Phys. Rev. Lett.\/}, vol.~84 (2000), pp.~4365--4368.

\bibitem{2001PhFl...13..561S}
{\sc R.~{Stieglitz} and U.~{M{\"u}ller}}.
\newblock {Experimental demonstration of a homogeneous two-scale dynamo}.
\newblock {\it \pof\/}, vol.~13 (2001), pp.~561--564.

\bibitem{2007PhRvL..98d4502M}
{\sc R.~{Monchaux}, et~al.}
\newblock {Generation of a Magnetic Field by Dynamo Action in a Turbulent Flow
  of Liquid Sodium}.
\newblock {\it Phys. Rev. Lett.\/}, vol.~98 (2007), no.~4, p.~044502.

\bibitem{2005EPJB...47..127A}
{\sc R.~{Avalos-Zu{\~n}iga} and F.~{Plunian}}.
\newblock {Influence of inner and outer walls electromagnetic properties on the
  onset of a stationary dynamo}.
\newblock {\it Eur. Phys. J. B\/}, vol.~47 (2005), pp.~127--135.

\bibitem{2005physics..11149S}
{\sc F.~{Stefani}, et~al.}
\newblock {Ambivalent effects of added layers on steady kinematic dynamos in
  cylindrical geometry: application to the VKS experiment}.
\newblock {\it Eur. J. Mech. B\/}, vol.~25 (2006), pp.~894--908.

\bibitem{2006JCoPh.218...44T}
{\sc R.~{Teyssier}, S.~{Fromang}, and E.~{Dormy}}.
\newblock {Kinematic dynamos using constrained transport with high order
  Godunov schemes and adaptive mesh refinement}.
\newblock {\it \jcp\/}, vol.~218 (2006), pp.~44--67.

\bibitem{2003guermond}
{\sc J.~L. {Guermond}, J.~{L\'eorat}, and C.~{Nore}}.
\newblock {A new Finite Element Method for magneto-dynamical problems:
  two-dimensional results}.
\newblock {\it Eur. J. Mech. B\/}, vol.~22 (2003), pp.~555--579.

\bibitem{2004PhPl...11.2838G}
{\sc A.~{Gailitis}, et~al.}
\newblock {Riga dynamo experiment and its theoretical background}.
\newblock {\it \pop\/}, vol.~11 (2004), pp.~2838--2843.

\bibitem{2006guermond_eccomas}
{\sc J.~L. {Guermond}, R.~{Laguerre}, J.~{L\'eorat}, and C.~{Nore}}.
\newblock {A finite element interior penalty method for MHD in heterogenous
  domains}.
\newblock In {\sc P.~{Wesseling}, E.~{O\~nate}, and J.~{P\'eriaux}}, editors,
  {\it European Conference on Computational Fluid Dynamics ECCOMAS CFD 2006\/}
  (TU Delft, Netherlands, 2006).

\bibitem{1997PhLA..226...75T}
{\sc A.~{Tilgner}}.
\newblock {A kinematic dynamo with a small scale velocity field}.
\newblock {\it Physics Letters A\/}, vol.~226 (1997), pp.~75--79.

\bibitem{2004JCoPh.196..102X}
{\sc M.~{Xu}, F.~{Stefani}, and G.~{Gerbeth}}.
\newblock {The integral equation method for a steady kinematic dynamo problem}.
\newblock {\it \jcp\/}, vol.~196 (2004), pp.~102--125.

\bibitem{2004PhRvE..70e6305X}
{\sc M.~{Xu}, F.~{Stefani}, and G.~{Gerbeth}}.
\newblock {Integral equation approach to time-dependent kinematic dynamos in
  finite domains}.
\newblock {\it \pre\/}, vol.~70 (2004), no.~5, p.~056305.

\bibitem{bem_brebbia}
{\sc C.~A. {Brebbia}, J.~C.~F. {Telles}, and L.~C. {Wrobel}}.
\newblock {\it Boundary Element Techniques\/} (Springer-Verlag, 1984).

\bibitem{2004JCoPh.197..540I}
{\sc A.~B. {Iskakov}, S.~{Descombes}, and E.~{Dormy}}.
\newblock {An integro-differential formulation for magnetic induction in
  bounded domains: boundary element-finite volume method}.
\newblock {\it \jcp\/}, vol.~197 (2004), pp.~540--554.

\bibitem{2001siam...22..1943H}
{\sc E.~{Haber} and U.~M. {Ascher}}.
\newblock {Fast Finite Volume Simulation of 3d electromagnetic problems with
  highly discontinuous coefficients}.
\newblock {\it SIAM J. Sci. Comput.\/}, vol.~22 (2001), pp.~1943--1961.

\bibitem{1992ApJS...80..753S}
{\sc J.~M. {Stone} and M.~L. {Norman}}.
\newblock {ZEUS-2D: A radiation magnetohydrodynamics code for astrophysical
  flows in two space dimensions. I - The hydrodynamic algorithms and tests.}
\newblock {\it \apjs\/}, vol.~80 (1992), pp.~753--790.

\bibitem{1992ApJS...80..791S}
{\sc J.~M. {Stone} and M.~L. {Norman}}.
\newblock {ZEUS-2D: A Radiation Magnetohydrodynamics Code for Astrophysical
  Flows in Two Space Dimensions. II. The Magnetohydrodynamic Algorithms and
  Tests}.
\newblock {\it \apjs\/}, vol.~80 (1992), p.~791.

\bibitem{2002JCoPh.183..165C}
{\sc G.~S. {Constantinescu} and S.~K. {Lele}}.
\newblock {A Highly Accurate Technique for the Treatment of Flow Equations at
  the Polar Axis in Cylindrical Coordinates Using Series Expansions}.
\newblock {\it \jcp\/}, vol.~183 (2002), pp.~165--186.

\bibitem{1987__telles}
{\sc J.~Telles}.
\newblock {A self-adaptive co-ordinate transformation for efficient numerical
  evaluation of general boundary element integrals}.
\newblock {\it Int. J. Num. Meth. Eng.\/}, vol.~24 (1987), no.~5, pp.~959--973.

\bibitem{2004phfl}
{\sc L.~{Mari{\'e}}, C.~{Normand}, and F.~{Daviaud}}.
\newblock {Galerkin analysis of kinematic dynamos in the von K{\'a}rm{\'a}n
  geometry}.
\newblock {\it Phys. Fluids\/}, vol.~18 (2004), pp.~017102--+.

\bibitem{2005PhFl...17k7104R}
{\sc F.~{Ravelet}, A.~{Chiffaudel}, F.~{Daviaud}, and J.~{L{\'e}orat}}.
\newblock {Toward an experimental von K{\'a}rm{\'a}n dynamo: Numerical studies
  for an optimized design}.
\newblock {\it Phys. Fluids\/}, vol.~17 (2005), pp.~7104--+.

\end{thebibliography}

\lastpageno     % This command sends the number of the last page to 
                % the MHD headline. Please latex your file twice if
                % you have used it.

%%%%%%%%%%%%%  
%       Place your tables and figures here
%%%%%%%%%%%%%

\begin{table}
\begin{tabular}{|c|r|r|r|r|}
\hline
Mode $m$ &  DEA & IEA & FV-BEM & FV (VTF)\\
\hline
0 &    --- &    --- &  -7.94 &  -5.81\\
1 & -8.13 & -8.42 & -8.75 & -7.20\\
\hline\end{tabular}
\caption{Decay rates for the free decay of an initial random field in a
  cylindrical geometry for the dipole
  mode ($m=0$) and the first non-axisymmetric mode ($m=1, \propto
  \cos\varphi$) obtained from a differential equation approach (DEA), the
  integral equation approach (IEA, see \cite{2005physics..11149S} for a
  description of both methods and the results), the hybrid finite volume-boundary element
  method presented here and the finite volume approach with vanishing tangential
  field conditions on the boundary.}
\label{tab::decay_rates}
\end{table}


\end{document}